\documentstyle[epsfig,
]{aipproc}

\begin{document}


\newcommand{\EEG}{\rm e^+ e^-\rightarrow \gamma\gamma}
\newcommand{\EEGG}{\rm e^+ e^-\rightarrow \gamma\gamma(\gamma)}
\newcommand{\EEGGG}{\rm e^+ e^-\rightarrow \gamma\gamma\gamma}
\newcommand{\EEEEG}{\rm e^+ e^-\rightarrow e^+e^-(\gamma)}
\newcommand{\LAMP}{ \Lambda_{+}}
\newcommand{\LAMM}{ \Lambda_{-}}
\newcommand{\LAMS}{ \Lambda_{6}}
\newcommand{\LAMPP}{ \Lambda_{++}}
\newcommand{\LAMMM}{ \Lambda_{--}}
\newcommand{\DSDW}{ \sigma(\theta)}
\newcommand{\MESTAR}{ m_{{\rm e^{\ast}}}}
\newcommand{\EELL}{\rm e^+ e^-\rightarrow l^{+}l^{-}}
\newcommand{\lum}{\rm pb^{-1}}
\def\beq{\begin{equation}} 
\def\eeq{\end{equation}}
\def\berr{\begin{eqnarray}}
\def\eerr{\end{eqnarray}}
\def\epem{{\rm e^+ e^-}}


\title{QED Test at LEP200 Energies in the Reaction $\EEGG$}


\author{ Jiawei Zhao on the behalf of the L3 Collaboration }


\maketitle

\begin{abstract}
The measurements of the QED reaction $ \EEGG $ performed
with the L3 detector are used to search for new
physics phenomena beyond the Standard Model. No evidence for
these phenomena is found and new limits on
their parameters are set. First the reaction is used
to constrain a model of an excited electron
and second to study
contact interactions.
The total and differential cross sections for the process
$ \EEGG $, are measured at energies from
91 GeV to 202 GeV using the data collected with
the L3 detector from 1991 to 1999. The L3 data
set lower limits
on the mass of an excited electron $ \MESTAR > 402 $ GeV,
on the QED cutoff parameters $ \LAMP > 415 $ GeV,
$ \LAMM > 258 $ GeV and
on the  contact interaction energy scale $ \Lambda > 1687 $ GeV.
The last parameter limits the size of the
interaction area to $ R < 1.17\times 10^{-17} $ cm. Some limits on the
string and quantum gravity scales are also discussed.  \end{abstract}

\section*{Introduction}
This paper describes a study of the process $\EEGG$ using data
\cite{L3old} and preliminary data \cite{L3new} recorded
with the L3 detector at LEP at center--of--mass energies (CM) from
91~GeV to 202~GeV. These processes are dominated by QED even at high LEP
energies. Since the differential cross--section is well known from QED
\cite{qed}, any deviation from this expectation hints at non--standard
physics processes contributing to the photonic final states. Any non--QED 
effects within the general framework of effective Lagrangian theory
are expected to increase with CM \cite{ebol}. A comparison of the
measured photon angular distribution with the QED expectation has been used
to put limits on the QED cut--off parameters \cite{LITKE}, and non--standard
$\gamma e^+e^-$ and $\gamma\gamma e^+e^-$ couplings \cite{ebol}. The
obtained limit on the QED cut--off parameters put constraints on the
size of the interaction area, which can be connected to the size of
selfgravitating particlelike structure with de Sitter vacuum core \cite{qed2}. The other 
application of our results is strongly related to the study of low energy
effects of large extra dimensions \cite{gia} in the string theory
framework \cite{string}.

We report on the measurements of total and differential cross section of
the $\EEGG$ reaction between 1991 and 1999 \cite{L3old,L3new}.
Previous results have also been published by the other groups \cite{collab}.
   
\begin{figure}[b!] \center{
{\centering \begin{tabular}{cc}
\resizebox*{0.5\textwidth}{0.3\textheight}{\includegraphics{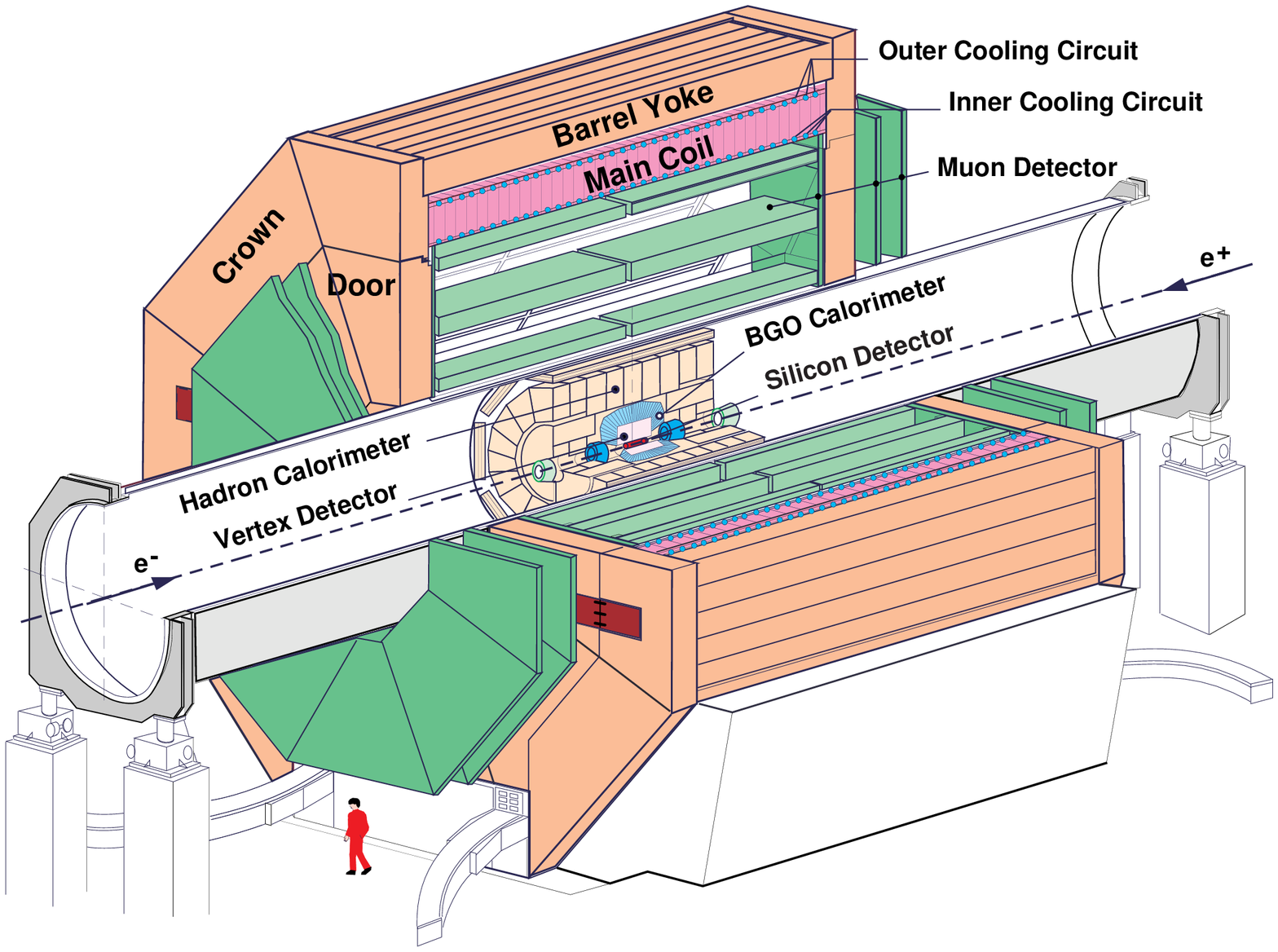}}&
\resizebox*{0.45\textwidth}{0.3\textheight}{\includegraphics{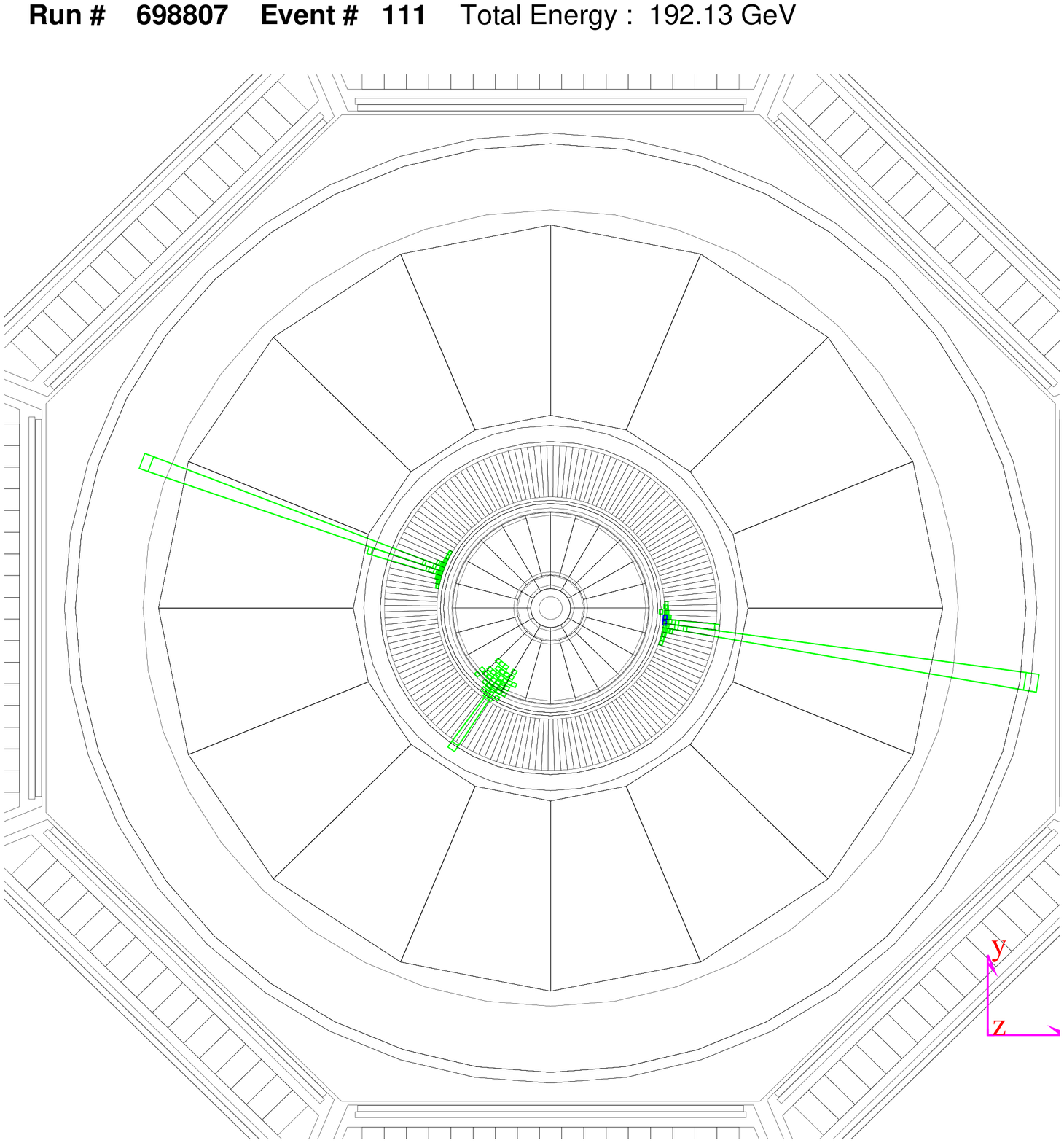}}\\
\end{tabular}\par}
}
\vspace{10pt}
\caption{The L3 detector (left side) and a typical $\EEGG$ event. The columns show the energy deposit in the BGO crystals.}
\label{detector}
\end{figure}

\section*{The L3 detector and event selection}
The L3 detector Fig.\ref{detector} is described in details in
\cite{detector}. The individual components of the detector are the vertex
silicon microstrip detector (SMD), the central tracking chamber (TEC)
working in a time expansion mode, the electromagnetic calorimeter composed
on bismuth germanium oxide crystals (BGO) with a barrel region
($42^{\circ}<\theta<138^{\circ}$) an two end-caps
($11^{\circ}<\theta<37^{\circ}$ and $143^{\circ}<\theta<169^{\circ}$), a
layer of scintillation counters used for time measurements, an uranium
hadronic calorimeter, and a high precision muon spectrometer. All
sub-detectors are located in a 12m diameter magnet which provides a
uniform field of 0.5 T parallel to the beam direction.  Forward BGO
calorimeter on either side of the detector measure the luminosity by
detecting small--angle Bhabha events.  Electrons and photons are measured
in the electromagnetic calorimeter with an energy resolution better than
2\% for particle energies above 1GeV. Fitting a shower shape to the
signals of the individual crystals of the electromagnetic bump results in
a spatial resolution of about 2 mm in front of the calorimeter
\cite{biland}. The discrimination of electrons and photons depends on the
track information of the central tracking chamber. This chamber is divided
into 24 outer sectors with 54 anode wires and 12 inner sectors with 8
anode wires. It detects charged particles reliably in the polar angle
range of $18^{\circ}<\theta <162^{\circ}$. The measurement of the
$(r,\phi)$ coordinates is performed with a single wire resolution of about
50 $\rm\mu m$.

For the event selection an ideal event of type $\EEGG$ has a
characteristic signature in the detector. Almost all of its energy is
deposited in the electromagnetic calorimeter (BGO) and there is no track
in the TEC Fig. \ref{detector}. 

To select an event there must be at least two photon candidates with 
polar angles $\theta_{\gamma}$ (angle with respective to the beam) between
$20^{\circ}$ and $160^{\circ}$  with an angular separation of more than
$60^{\circ}$ between two most energetic photons, and no other activity in
the detector. 

A shower in the electromagnetic calorimeter must show a profile
consistent with that of a photon and an energy above 2 GeV. To reject 
$\epem \rightarrow \nu \bar{\nu} \gamma \gamma$ and cosmic rays, we
require that the sum of the  energies of the photon candidates be larger
than $\sqrt{s}/2$. The remaining backgrounds  are $\epem \rightarrow \epem
(\gamma\gamma )$ and $\epem \rightarrow \tau^{+} \tau^{-} (\gamma\gamma )$
with charged particles in the beam pipe. These  contributions are
estimated from Monte Carlo (MC) simulations using  BHWIDE for Bhabha events and
KORALZ for  $\tau$ events, and are found to be negligible. The acceptance
is computed  applying the same analysis to a  sample of
$\epem\rightarrow\gamma\gamma (\gamma)$ events generated using an  order
$\alpha^{3}$ MC generator \cite{qed,L3old}  passed through the L3
simulation (see refs. in \cite{L3old}) and reconstruction  programs. As an
example, the selection efficiencies to detect at least
two photons inside the fiducial volume are represented in Table
\ref{table1} for four energies. The efficiency of the calorimetric energy 
trigger is estimated to be above 99.7 \% .

\begin{table}
\caption{The luminosity, selection efficiencies at the different CM's,  the  number of observed events and the total cross section.} \label{table1} 
\begin{tabular}{lcccc}
  \multicolumn{1}{c} {Energy (GeV)}& Luminosity ($\lum$)&
   \multicolumn{1}{c}{Efficiency (\%)} &
  \multicolumn{1}{c}{Event number} & 
\multicolumn{1}{c}
{$\sigma_{\EEGG}^{tot}$ (pb)} \\
\tableline
192 & 29 & 61 & 154 
& $\rm 8.74\pm 0.71\pm 0.17$ 
\\
196 & 82 & 61 & 427 
& $\rm 8.51\pm 0.41\pm 0.16$ 
\\
200 & 76  & 62 & 425 
& $\rm 9.13\pm 0.44\pm 0.17$ 
\\
202 & 37  & 62 & 163 
& $\rm 7.15\pm 0.56\pm 0.14$ 
\\
\end{tabular}
\end{table}

\begin{figure} 
\centerline{\epsfig{file=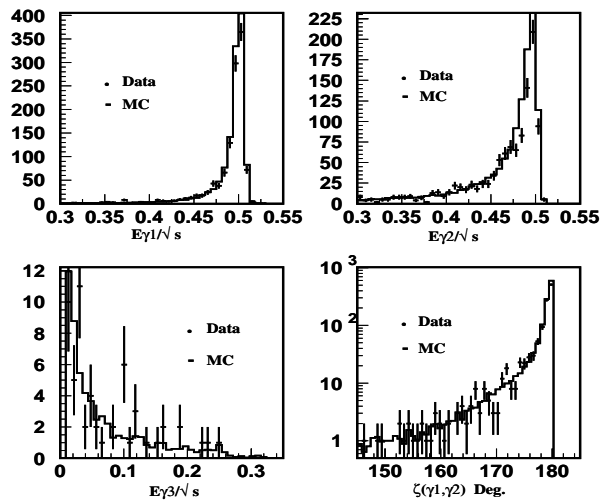,
height=3.0in,width=3.5in}}
\caption{The photon energy $E_{\gamma i}$ ($i=1,2,3$) and the collinearity angle distributions are compared with the MC expectation.} \label{eff}
\end{figure}

\section*{analyzes of $\EEGG$ events}

Applying this selection cuts the number of observed events,
classified according to the number of isolated photons within the range
$20^{\circ}<\theta_{\gamma}<160^{\circ}$, as presented in Table \ref{table1}
at the four different CM energies. No events with 5 or more photons have
been observed. 

To check the consistency of data and MC, the photons in the events are
sorted with respect to their energy $E_{\gamma 1}>E_{\gamma
2}>E_{\gamma 3}$ and compared to MC. This is shown in Fig. \ref{eff}
together with the  distribution of the collinearity angle of first two
photons, normalized to 200 GeV.  Data and MC are in good
agreement for all energies. 

\begin{figure} 
\center{
{\centering \begin{tabular}{cc}
\resizebox*{0.5\textwidth}{0.3\textheight}{\includegraphics{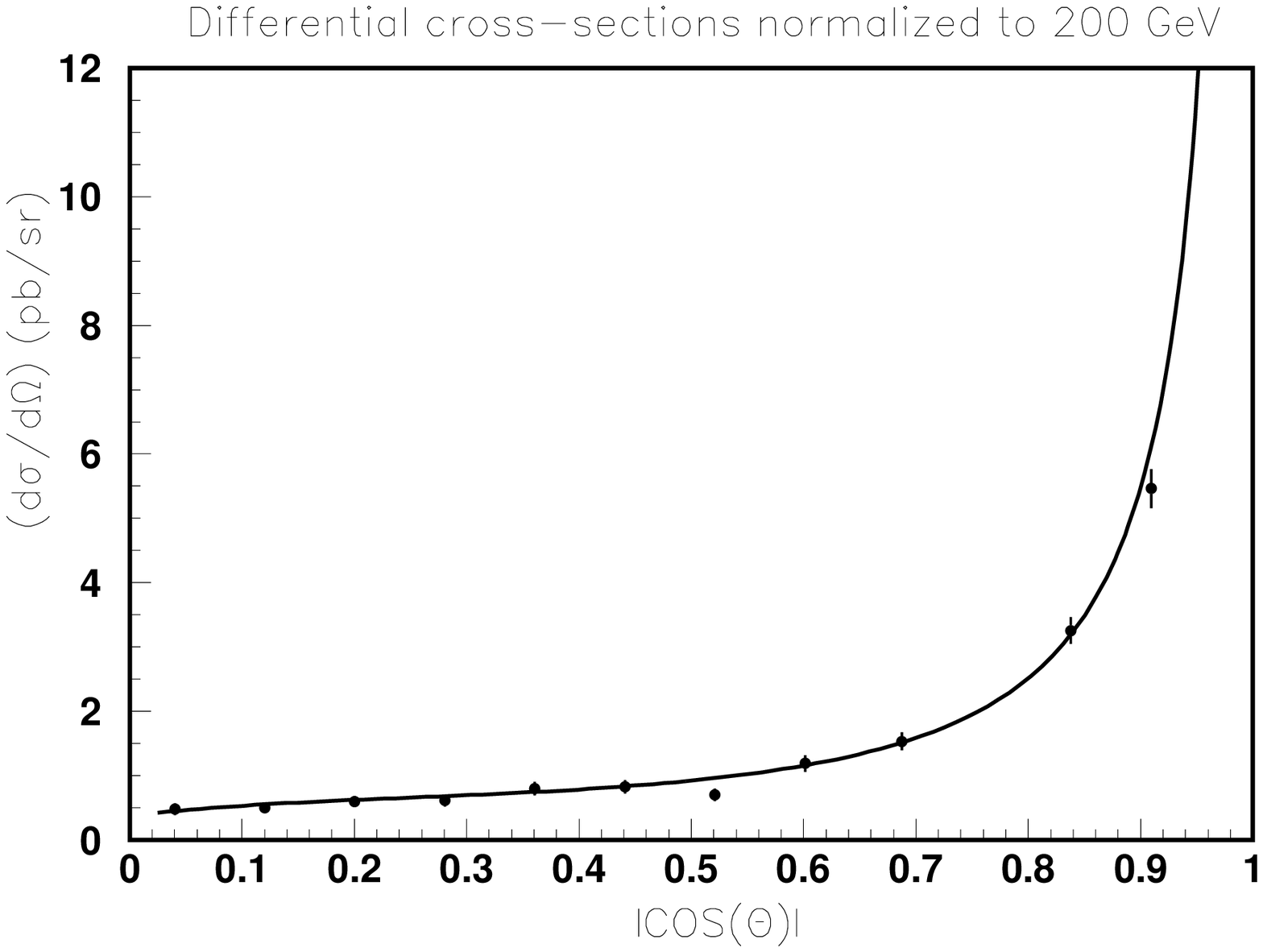}}&
\resizebox*{0.5\textwidth}{0.3\textheight}{\includegraphics{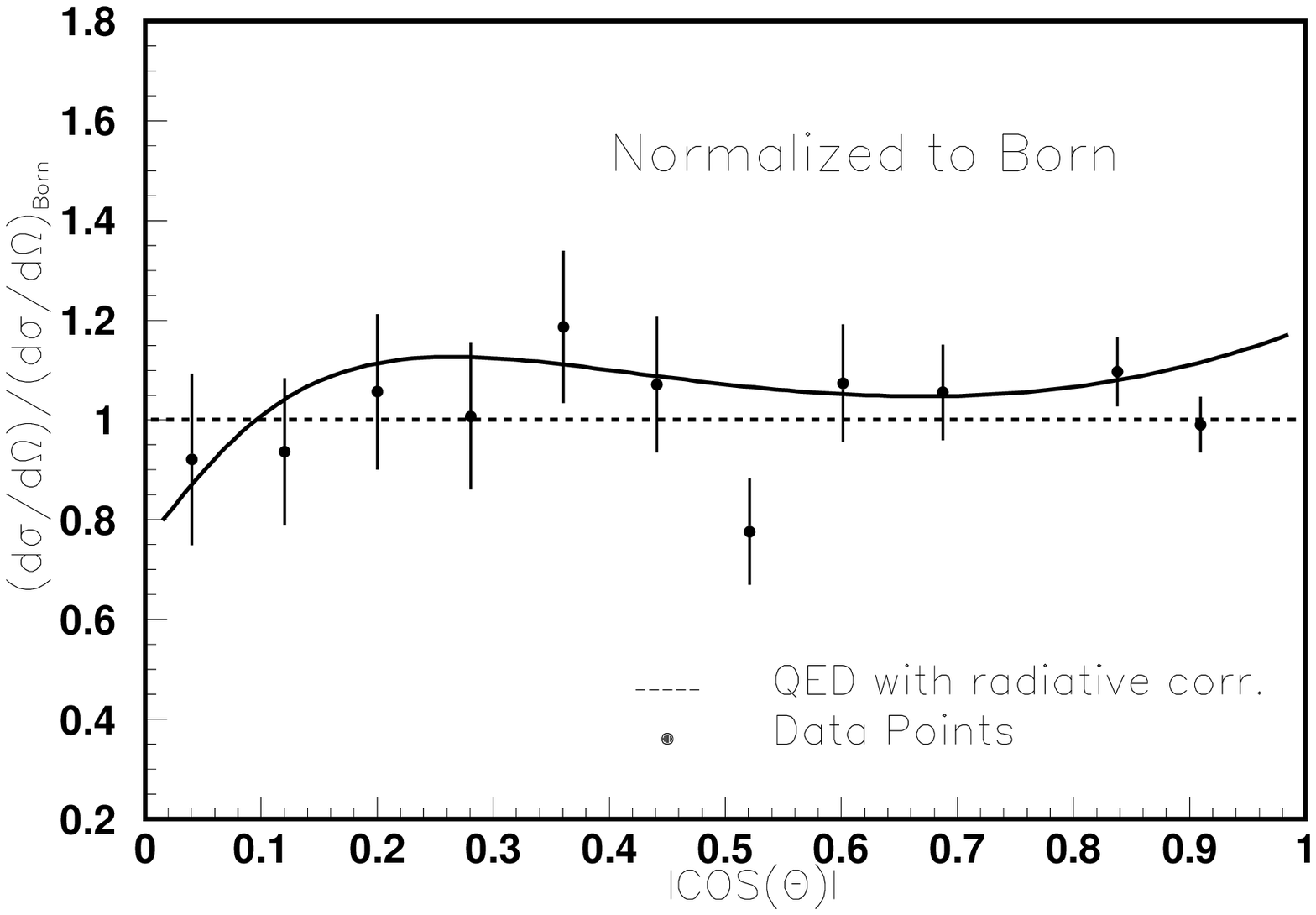}}\\
\end{tabular}\par}
}
\caption{The measured angular distribution for the process $\EEGG$ normalized 
to 200GeV. The points show the experimental data. The solid curve on the
left side corresponds to the QED ${\cal O}(\alpha^3)$ prediction. The
ratio  $(d\sigma /d\Omega )/(d\sigma /d\Omega )_{Born}$ is displayed on the
right side.} \label{cros} \end{figure}

The knowledge of the number of events, luminosity and efficiencies allows us to
calculate the differential cross section, which is shown in Fig.\ref{cros}
(left side) and normalized to the Born level (right side). The polar
angle $\Theta$ of event is defined as  $\cos\Theta
=\frac{1}{2}(|\cos\theta_1|+|\cos\theta_2|)$, where $\theta_1$ and
$\theta_2$ are the polar angles of the two most energetic photons in the
event. Radiative corrections are quantified in Fig. \ref{cros} (right
side) as the difference between the differential cross section up to
${\cal O}(\alpha^3)$ (solid line) and the Born level (broken line).
Excellent agreement between the data and QED predictions is observed 
up to ${\cal O}(\alpha^3)$.

\begin{figure} [b!]
\centerline{\epsfig{file=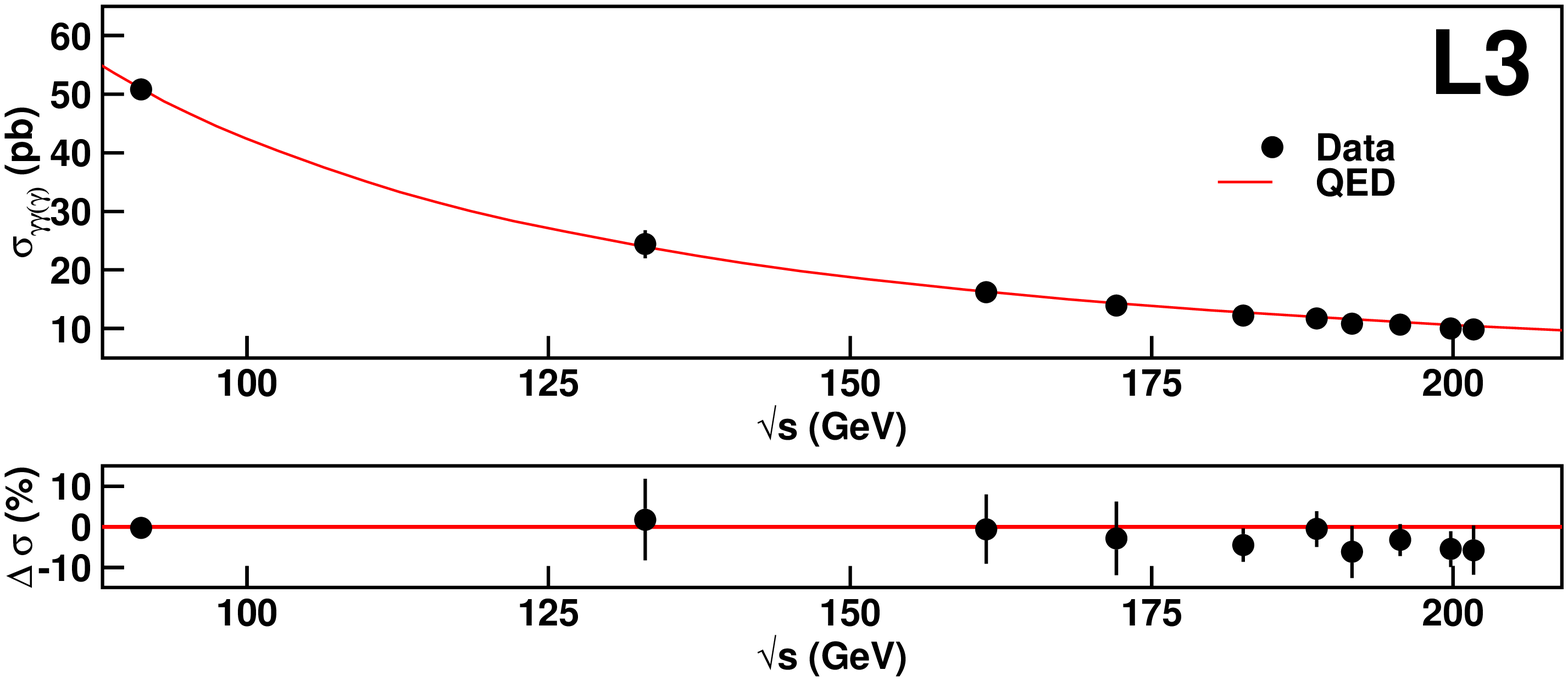,
height=3.1in,width=3.5in}}
\caption{Total measured cross-section (points) compared with QED
prediction up to ${\cal O}(\alpha^3)$ and $\Delta\sigma (\% )$ as a
function of CM energy.} \label{simat} \end{figure}
 
The observed number of events in the fiducial region
$20^{\circ}<\theta<160^{\circ}$ corresponds to the total cross sections,
which are presented in Table. \ref{table1}. The first error is statistical
and the second one is systematic. The statistical error dominates in the
measurement of the cross section at every energy. It has been evaluated by
varying the selection cuts and taking into account the finite MC statistic.

A second independent data analysis, with similar cuts,  from 90GeV to
202GeV was performed measuring the differential and total
cross section of the $\EEGG$ reaction. Both analysis are agreed.
Fig.\ref{simat} shows the total cross section of this analysis compared
to the QED predictions up to ${\cal O}(\alpha^3)$ level. 

\section*{Limits on deviations from QED}

In the case of electromagnetic interaction
the process $ \EEGG $ is ideal to test the QED because
it is not interfered by the $ Z^{o} $ decay. This
reaction proceeds via the exchange of a virtual electron
in the t -- and u -- channels, while the s -- channel is
forbidden due to the angular momentum conservation.

The differential cross section for the process $\EEG$ in the relativistic
limit of lower order QED is given by \cite{born}: \beq
\label{born}
\left(\frac{d\sigma}{d\Omega}\right)_{Born}
=\frac{\alpha^2}{s}\cdot\frac{1+\cos^2\theta}{1-\cos^2\theta} \eeq      
where $\alpha$ is the electromagnetic coupling constant and $\theta$ is the
polar angle of one of the photons. Since two photons can not be
distinguished the event angle is defined as positive.  A possible
deviation from the QED cross section for Bhabha and M\"oller scattering are
parameterized in terms of cutoff parameters $\Lambda$. Such parameters
correspond to a short range exponential term added to the Coulomb potential.
This ansatz together with  ${\cal O}(\alpha^3)$ radiative corrections
leads to a modification of the photon angular distribution
(\ref{born}) \beq \label{borncor}
\left(\frac{d\sigma}{d\Omega}\right)_{QED+DEV}=
\left(\frac{d\sigma}{d\Omega}\right)_{O(\alpha^3)}(1+\delta_{DEV}), \eeq

In the present paper we use the  agreement between the data and the QED
predictions to constrain
the model of an excited
electron of mass $ m_{e^{*}} $ which
replaces the virtual
electron in the QED process \cite{LITKE}, as well as to constrain
the model with deviation from QED arising from an
effective interaction with non--standard
$ e^{+} e^{-} \gamma $ couplings and
$ e^{+} e^{-} \gamma \gamma $ contact terms \cite{ebol}.

The heavy excited electron couples
to an electron and a photon via magnetic interaction
with an effective gauge invariant Lagrangian of \cite{LITKE}.

\begin{equation}
\label{alleqs1}
{\mathcal L}_{\text{excited}}=\frac{e\lambda}{2m_{e^{*}}}
\overline{\psi_{e^{*}}}\sigma_{\mu\nu}\psi_{e}F^{\mu\nu}
\label{eq.1}
\end{equation}

In this equation $ \lambda $ is the coupling constant,
$ F^{\mu\nu} $ the strength
electromagnetic field tensor,
$ \psi_{e^{*}} $ and $ \psi_{e} $ are the wave functions
of the heavy electron and the ordinary electron respectively.

In the case of effective contact interaction with non--standard
coupling a cut--off parameter $ \Lambda $
is introduced to describe the scale of the interaction
with the following Lagrangian \cite{ebol}

\begin{equation}
\label{alleqs2}
{\mathcal L}_{\text{contact}}=i\overline{\psi_{e}}
\gamma_{\mu}(D_{\nu}\psi_{e})
\left(\frac{\sqrt{4\pi}}{\Lambda^{2}_{6}}F^{\mu\nu}+
\frac{\sqrt{4\pi}}{\tilde{\Lambda}^{2}_{6}}\tilde{F}^{\mu\nu}\right)
\label{eq.2}
\end{equation}

The effective Lagrangian chosen in this case has an operator
of dimension 6, the wave function of the electron
is $ \psi_{e} $, the QED covariant derivative is $ D_{\nu} $,
the tilde on $ \tilde{\Lambda}_{6} $ and $ \tilde{F}^{\mu\nu} $
stands for dual.

In the case of the excited electron, if the
CM energy $\sqrt{s}$ satisfies
the condition $ s/m^{2}_{e^{*}}\ll 1$, then $\delta_{DEV}$
reads

\begin{equation}
\label{alleqs4}
\delta_{DEV}=\pm s^{2}/2 ( 1/\Lambda^{4}_{\pm})
(1-\cos^{2}\theta)
\label{eq.4}
\end{equation}

In this approximation, the parameters $\Lambda_{\pm} $ are the
QED cut--off parameters with
$ \Lambda^{2}_{+}=m^{2}_{e^{*}}/\lambda $. In the case of an arbitrary
value of $\sqrt{s}$ the full equation of ref.\cite{LITKE} is
used to calculate
$ \delta_{DEV} = f(m_{e^{*}}) $.

\begin{figure} 
{\centering \begin{tabular}{cc}
\resizebox*{0.5\textwidth}{0.3\textheight}{\includegraphics{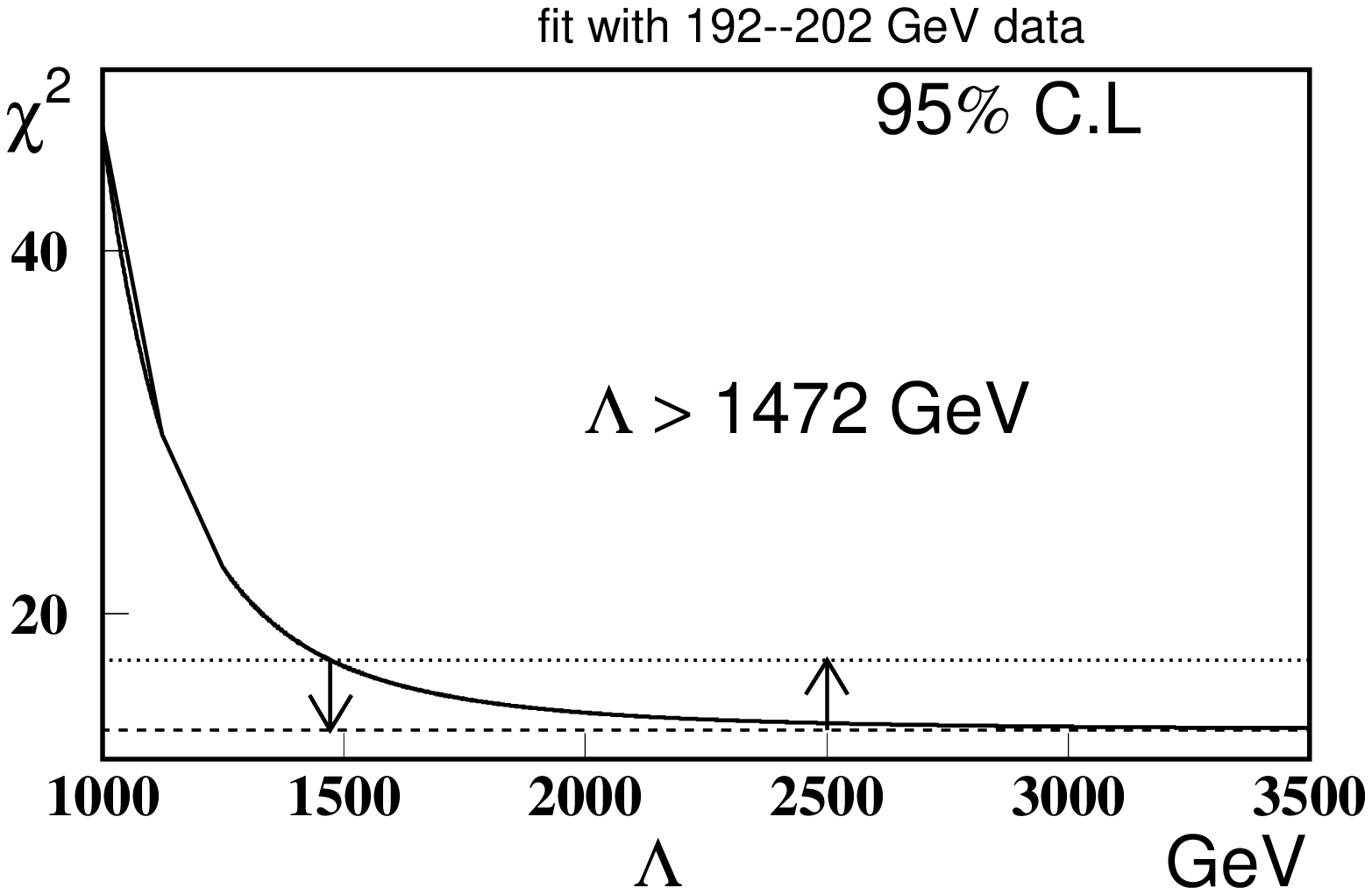}}&
\resizebox*{0.5\textwidth}{0.3\textheight}{\includegraphics{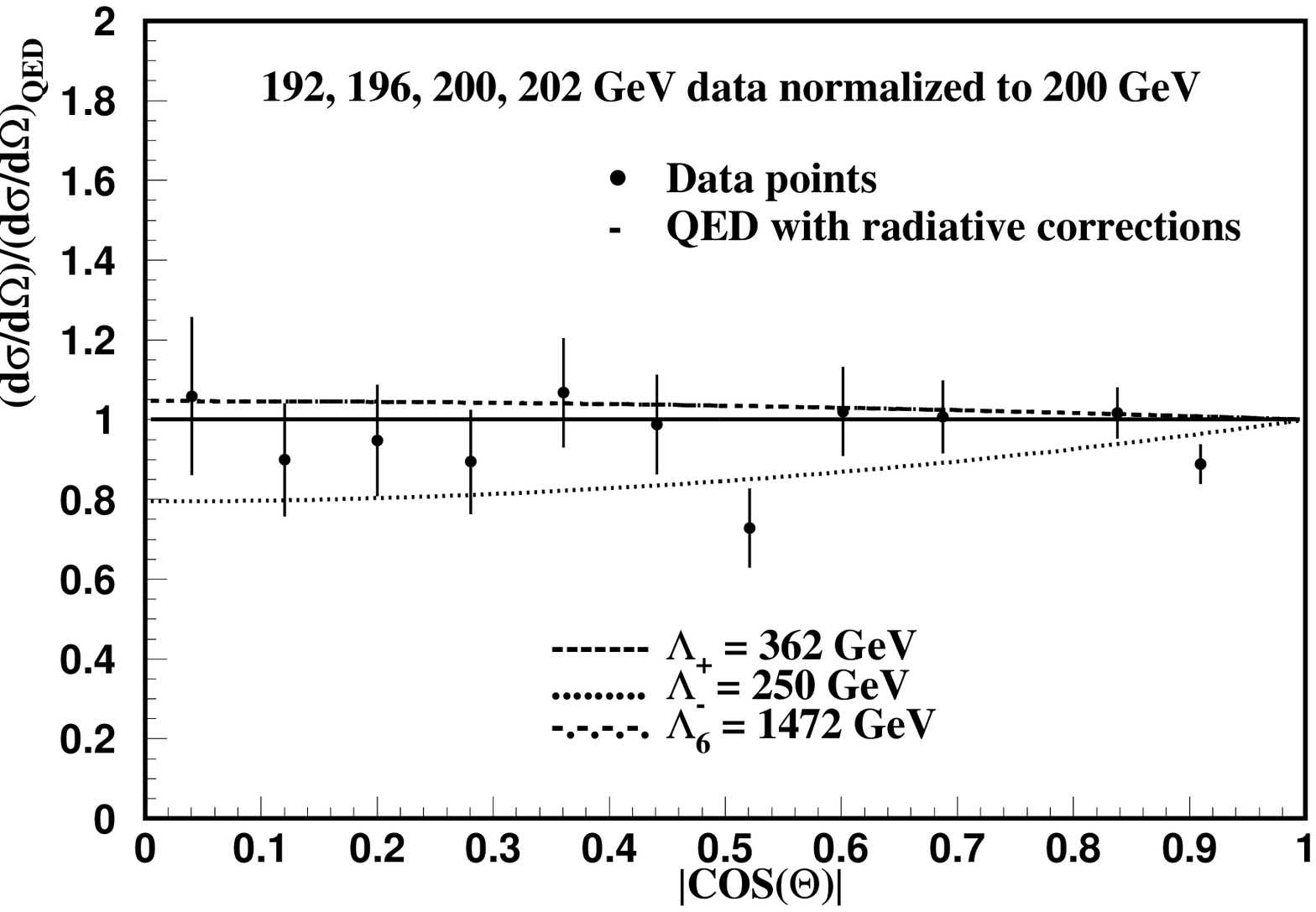}}\\
\end{tabular}\par}
\caption{The behavior of $\chi^2$ as a function from $\Lambda$. The comparison of the measured differential cross section with the QED predictions including the deviations of the parameters $\Lambda_{\pm}$ and $\Lambda_6$ as a function of $|\cos\Theta|$. The cross section is normalized to the radiatively ${\cal O}(\alpha^3)$ corrected QED cross section (right side).}
\label{tup}
\end{figure}

In the case of contact interaction $\delta_{DEV}$ reads as
\begin{equation}
\delta_{DEV}=s^{2}/(2\alpha)(1/\Lambda^{4}_{6} + 1/\tilde{\Lambda}^{4}_{6})
(1-\cos^{2}\theta)
\label{eq.5}
\end{equation}
At the CM under consideration, the $\chi^2$ fit
 gives for the excited electron~$\MESTAR >343$~GeV with
the QED cut--off parameters $ \LAMP > 362 $ GeV and
$ \LAMM > 250 $ GeV and for non--point like coupling $\Lambda >1472$GeV
at 95\% CL (we set $ \Lambda_{6}=\tilde{\Lambda}_{6}=\Lambda $). The $\chi^2$ as
a function of $\Lambda$ is shown in Fig. \ref{tup} (left side) together
with differential cross section normalized to the QED one at the ${\cal
O}(\alpha )^3$ level (right side). The most stringent limit we get from a
global $\chi^2$ fit from the results at the CM's 91.2GeV, 133GeV, 161GeV,
172GeV, 183GeV, 189GeV and 200GeV. These limits read $\MESTAR
>402$GeV, $ \LAMP > 415 $ GeV, $ \LAMM > 258 $ GeV and $\Lambda >1687$GeV
at 95\% CL. 
These scales allow to estimate the upper limits on the characteristic 
size of the interaction region related to the interactions
(\ref{alleqs2}) and  (\ref{alleqs1})  respectively. The limit on the
contact interaction establishes the characteristic limit  of the QED size
interaction region to $R\sim{\hbar / (m_{e^*}c) < } 1.17 \times 10^{-17} $
cm. The behavior of $ \chi^{2} $ as a function of $ \Lambda $ shows no
minimum indicating that the size of the interaction region
must be smaller than $R$. 
We conclude that the most stringent upper limit on scale of the
non-point like behavior of electron comes from the contact interaction
term.   

We found also that all experimental limits on sizes of fundamental particles
are smaller than their Compton wavelengths. 
This fact has been applied to 
estimate the lower limits on particle sizes in 
the model of a fundamental particle as 
a selfgravitating structure with de Sitter vacuum core \cite{qed2}.

Our results could also be applied to study  TeV scale quantum gravity \cite{gia}. It is possible to suggest that the fundamental scale of gravitational interaction $M$ is as low as TeV \cite{gia}, whereas the observed weakness of the Newtonian coupling constant $G_N\sim M_{Pl}^{-2}$ is due to the existence of N large ($\ell\gg\rm TeV^{-1}$) extra dimensions into which the gravitational flux can spread out.  At the distances larger than the typical size of these extra dimensions the gravity goes to its standard Einstein form, and  the usual Newtonian low can be recovered via the relation $M_{Pl}=M^{N+2}\ell^N$ \cite{gia} between Plank scale and scale $M$. It means that, such kind of quantum gravity becomes strong at the energies $M$, where presumably all the interactions must unify, without any hierarchy problem. The phenomenological implications of large extra dimensions is
concentrated on the effects of real and virtual graviton emission. The
basic assumption is, that gravitons can propagate in extra dimensions \cite{gia}. The
quantum states of such gravitons are characterized by quantized momentum
in the large extra dimensions. The only known
framework that allows a selfconsistent description of quantum gravity
is string theory. As an essential part of the structure of string theory
\cite{stringtheory} is that the gravitons and the particles of standard
model must have an extended structure. This means that, there will be
additional modifications of standard model amplitudes due to string
excitations which can compete with or even overwhelm the modifications due
to graviton exchange \cite{string}. An important effects of simple model
of string theory with large extra dimensions \cite{string} come from the
exchange of string Regge (SR) excitations of standard particles. In
standard model scattering processes, contact interactions due to SR
exchange produce their own characteristic effects in differential cross
section, and these effects typically dominate the effects due to Kaluza --
Klein (KK) exchange \cite{gia}. The SR excitation effects can be visible
as contact interactions \cite{string} well below the string scale $M_S$.
The deviation from the standard model, we investigated, has been performed
in the terms of Drell's parameterization (\ref{alleqs4}). Actually, this
parameterization is applicable to any beyond standard model at short
distances. Thus, from the comparison of string theory result \cite{string}
to the (\ref{alleqs4}) the following identification can be deduced: $
\Lambda_+=\left(\frac{12}{\pi^2}\right)^{1/4}M_S $. Our result
$\Lambda_+>415$~GeV corresponds to $M_S>396$~GeV at 95\% C.L. Using the
connection between string scale and quantum gravity scale $M$
\cite{string} we find $M>1188$~GeV.


\begin{references}
\bibitem{L3old} L3 Collaboration, Acciarri, M., et al.,
                {\it Phys.\ Lett.} {\bf B384}, 323 (1996);

                L3 Collaboration, Acciarri, M., et al.,
                {\it Phys.\ Lett.} {\bf B413}, 159 (1997);

                L3 Collaboration, Acciarri, M., et al.,
                {\it Phys.\ Lett.} {\bf B475}, 198 (2000).
 \bibitem{L3new} L3 Collab., L3 Internal Note 2627 ( 2000 ),
                 unpublished.,

                 L3 Collab., Acciarri, M., et al.,
                 {\it Proceedings of
                 International Europhysics Conference on High Energy
                 Physics Budapest, Hungary, 12-18 July 2001,}
                 Journal of high energy physics, PRHEP-hep2001/248
\bibitem{qed}Berends, F. A., {\it Nucl.\ Phys.} {\bf B186}, 22 (1981).
\bibitem{ebol} Eboli, O. J. P., Natale, A. A., and
               Novaes, S. F., {\it Phys.\ Lett.} {\bf B271}, 274 (1991).
\bibitem{LITKE}
               A. Litke, Harvard Univ., {\it Ph.D\ Thesis}, unpublished (1970).
\bibitem{qed2} Dymnikova, I., Ulbricht, J., and Zhao, J., hep-ph/0109138.
\bibitem{gia} Arkani--Hamed, N., Dimopoulos, S., and Dvali, G., {\it Phys.\ Lett.} 
{\bf B429}, 263 (1998); Antoniadis, I., Arkani--Hamed, N.,
Dimopoulos, S., and Dvali, G., {\it Phys.\ Lett.} {\bf 436}, 257 (1998).
\bibitem{string}Cullen, S., Perelstein, M., and Peskin, M. E., {\it Phys.\ Rev.} 
{\bf D62}, 055012 (2000).
\bibitem{collab} DELPHI Collaboration, Abreu, P., et al,
                 {\it Phys.\ Lett.} submitted; \\
                 ALEPH Collaboration, Barate, R., et al, {\it Phys.\ Lett.}
                 {\bf B429}, 201 (1998);       \\
                  OPAL Collaboration,
                 Abbiendi,~K., et al, {\it Phys.\ Lett.} {\bf B465}, 303 (1999).
\bibitem{detector}L3 Collaboration, Adeva, B., et al,
                 {\it Nucl.\ Inst.\ Meth.} {\bf A289}, 35 (1990).
\bibitem{biland}L3 Collaboration, Adrian, O., et al,
                {\it Phys.\ Rep.} {\bf 236}, 1 (1990).
\bibitem{born}Harris, I., and Brown, L. M., {\it Phys.\ Rev.}
             {\bf 105}, 1656 (1957).
\bibitem{stringtheory} Polchinsky, J., {\it ``String\ Theory''},
                       Camrige University Press (1998).

\end{references}
\end{document}